\documentclass[rnote]{aa} 
\usepackage{graphicx}
\usepackage[]{natbib}
\usepackage{txfonts}

\newcommand{\meta}{[Fe/H]}
\newcommand{\T}{$T_{\rm eff}$}
\newcommand{\g}{log($g$)}
\newcommand{\Vmicro}{$V_{\rm micro}$}
\newcommand{\mic}{$\mu {\rm m}$}

\usepackage{color}

\begin{document}

   \title{Revisited fluorine abundances in the globular cluster M~22 (NGC~6656)}

   \author{P. de Laverny\inst{1}
          \and
          A. Recio-Blanco\inst{1}
          }

   \institute{Laboratoire Lagrange (UMR7293), Universit\'e de Nice Sophia Antipolis, CNRS, Observatoire de la C\^ote d'Azur, BP 4229,
 F-06304 Nice cedex 4, France\\
              \email{laverny@oca.eu}
             }

   \date{Received 2013; accepted}

   \abstract
    {}   
    {Fluorine is a fairly good tracer of formation histories of multiple stellar populations in globular clusters as already revealed by several studies.
Large variations in fluorine abundance in red giant stars of the globular cluster M~22  have been recently reported by
two different groups. Futhermore, one of these studies claims that the abundance of fluorine is anti-correlated
with sodium abundances in this cluster, leading to strong conclusions on the chemical history of M~22. To validate this important finding,
we re-examine the F abundance determinations of some of the previously studied stars.
}
    {We have reanalysed some high-resolution VLT/CRIRES spectra of RGB stars found in M~22 in order to re-estimate their fluorine abundance from the 
spectral synthesis of the HF line at 2.336$\mu$m.}
    {Unlike what has been previously estimated, we show that only upper limits or doubtful fluorine abundances with large uncertainties
in M~22 RGB stars can be derived. This is caused 
by an incorrect identification of continuum fluctuations as the HF signature combined with a wrong correction
of the stellar radial velocity. Such continuum fluctuations could be the consequences of
telluric residuals that are still 
present in the analysed spectra. Therefore, no definitive conclusions on the 
chemical pollution caused by the
M~22 first stellar generation can presently be drawn from the fluorine content of this cluster.}
  {} 


   \maketitle
%

\section{Introduction}

Fluorine is a chemical species of special interest in studies devoted
to the evolution history of stellar populations for several reasons.
For instance,
(i) the nucleosynthetic origin of fluorine in the Universe is still debated with
two main ranges of stellar masses proposed for its production site, and
(ii) fluorine abundances may help disentangle the complex formation history of the
globular clusters (GCs)
by looking for the chemical pollution signatures of their first stellar population.
Indeed, over the last decade, 
several studies devoted to the abundance of fluorine in
different types of galactic and extragalactic stellar populations have shed new light on
their chemical history \citep[see][for a more detailed
description]{Recio12, deLaverny13}.

In particular, the abundances of fluorine on the surface of 
some red giant branch stars (RGBs) of a few globular clusters have already been estimated. Most of these targeted GCs are metal-rich or have an intermediate
metallicity (typically, \meta $<$ -1.2). For more metal-poor clusters,
only upper limits of the fluorine content of NGC~6397 ([Fe/H] = -1.99)
and M~30 ([Fe/H] = -2.33) have been reported up to now \citep{deLaverny13}. Unfortunately,
these too high upper limits avoid any derivation of any conclusive information
on the chemical history of these clusters. It has also been shown that these 
F abundance upper limits in such metal-poor GCs are probably caused by 
inherent limitations of high-resolution $K$-band spectra collected
with current telescopes and spectrographs.

In that figure, the only exception of fluorine abundance measurements
(and not only upper limits) 
in metal-poor GCs is M~22 (NGC~6656) with a mean metallicity [Fe/H]$\sim$-1.7~dex.
Recently, \citet{Alves-Brito12} have indeed reported [F/Fe] ratio estimates in 
five RGB stars and upper limits in two other stars of M~22.
Then, \citet[][DLL13 hereafter]{Dorazi12} partly revised these abundances:  their sample
of six M~22 RGBs contains four stars already studied by \citet{Alves-Brito12}.
DLL13 show that the telluric substraction  
may partly explain their different estimates of F-abundances
with respect to the previous study. These new abundance estimations 
allowed them
to derive constraints on the masses of the first-generation stars
that could have chemically polluted M~22.

However, our previous analysis of the fluorine content in metal-poor GCs
\citep{deLaverny13} has shown that F-abundances cannot be derived
well in such clusters. In this research note and in to validate the F-abundances
previously estimated in M~22, we thus
re-examine some of the spectra collected by DLL13.
We show that the fluorine content of M~22 RGB members cannot be safely
estimated (see Sect.~\ref{Sect:Fluor}). This leads us to conclude 
in Sect.~\ref{Sect:conclu} that the properties of
the M~22 first-generation stars (as the range of masses of the main polluters) 
cannot currently be drawn from the fluorine content of this cluster.

\section{Analysis of M~22 RGB CRIRES spectra around the fluorine signature}
\label{Sect:Fluor}

We selected three RGB stars of the sample of DLL13 and re-analysed their spectra.
We recall that these spectra were collected with the VLT/CRIRES instrument (see DLL13, for a description of the observations, their reduction
and telluric features substraction). We point out that we re-analysed 
exactly the same spectra as DLL13 who performed the cleaning from the
telluric contamination. In this note, we indeed only focus on the re-estimation of
the fluorine abundances from these already reduced and cleaned spectra.

The selected targets are the stars {\it \#III-15, \#C}, and {\it \#III-52}. The first one has been estimated by DLL13  to 
be the most fluorine-poor
([F/Fe]=-1.0~dex) of their sample  of six M~22 RGBs, whereas the other two are proposed as 
quite enriched in F (with [F/Fe]=-0.8 and -0.6~dex, respectively). Actually, it
is claimed that the star {\it \#III-52} would be the most enriched in fluorine found so far in M~22. 

The new fluorine abundances were estimated by adopting exactly the same methodology as in \citet{deLaverny13}, which is actually identical to all our previous works
devoted to F abundance determinations \citep[see, for instance, our first paper in this series,]
[]{Abia09}. We only recall that these
abundance determinations are based on the synthetic modelling
of the unblended HF(1-0)~R9 line at 2.3358~\mic . 
We also adopted the same atmospheric parameters \T , \g , \meta, and \Vmicro \ as DLL13.
The synthetic spectra were then broadened by convolution
with a Gaussian profile of FWHM~=~9~km/s to match the observed line widths.

We show in Fig.~1 the synthetic fits of the three selected spectra
around the HF transition, assuming
three different fluorine abundances spanning at least 1.3~dex.
The largest of these abundances could actually be considered as an upper limit of A(F). From this Fig.~1, 
it can be clearly seen that no reliable fluorine
abundances can be derived from these spectra.
Indeed, if the stellar radial velocity is corrected well to match all the 
strong CO-lines (left panels of Fig.~1), no line feature can easily be identified at 
the right-hand HF-line position (see the plots in the right-hand panels of Fig.~1, 
showing a zoom around the HF-line). 
It can indeed be seen that
large flux fluctuations are present around the HF line, whereas
no features are predicted by the synthetic spectra
in these almost perfectly line-free (continuum) regions of very metal-poor stars. 
Furthermore, these fluctuations
have the same amplitude (a few percent) as the expected HF line in these stars.
They could thus mimic the presence of the HF transition if examined
over too narrow a spectral domain (see, for instance, Fig.~2 of DLL13)
but, then, losing the right position for the CO lines.
Therefore, for two of the stars shown in Fig.~1 ({\it \#III-15} and {\it \#III-52}),
 only an upper
limit of their fluorine content could be estimated safely (A(F) $\le$ 2.3 and 2.65~dex,
respectively). For the RGB star {\it \#C}, if the feature seen close to the HF-line
position were real (although not exactly at the right wavelength), 
it could only be estimated that its F-content were
found in a wide range [2.0 - 2.6].
However, we again point out that we cannot exclude that no HF-line 
signature at all is present
in these three spectra characterized by continuum flux fluctuations
at almost all wavelengths and with similar amplitudes as the supposed HF feature.

On the other hand, we think that these flux variations could be caused
by the telluric contamination that
cannot be perfectly cleaned in such spectra. Even with an optimized procedure
to remove the telluric absorption lines,
unfortunately for a cluster like M~22, the 
main telluric feature that contaminates the spectra in that spectral range
has a wing that can still be strong around the HF position. Owing to possible limitations on the stability of CRIRES and/or the 
earth atmosphere, subtracting the two spectra
(the RGB target and the telluric standard, i.e. a hot star) could artificially produce
residuals with almost the same amplitude (1\% to 2\%)
and position as the HF-transition. Then, this could lead to possible
incorrect identifications of the faintest stellar features.
Actually, this possible explanation is reinforced by examining  Fig.~2 where we compare
the spectra of the star {\it \#III-52} with and without telluric correction.
We recall that this star is the F-richest of the DDL sample.
It can be seen that the telluric absorption lines are large in this spectral domain 
and that all the continuum fluctuations seen in the cleaned 
{\it \#III-52} spectrum are located at the telluric absorptions.
In particular, because of its radial velocity, the studied HF transition
is located at a telluric line that absorbs $\sim$7\%
of the continuum flux. After telluric cleaning, several continuum fluctuations
are still present. In particular, a continuum fluctuation
can be seen with an amplitude as large as the expected HF line ($\sim$1\%)
but shifted by two to three pixels with respect to the right HF-line wavelength.
This artefact could thus mimic an HF-line if the stellar radial velocity is
not perfectly estimated with the neighbouring CO lines (see Fig.~1), and then corrected.
Furthermore, this Fig.~2 can be compared to the Fig.~7 of DLL13 for the star {\it \#III-15} that has been
proposed to be the most fluorine-poor of M~22 by these authors. It can be seen in their figure (lower
panel) that the telluric contribution is almost null at the HF-line position, because
of the {\it \#III-15} radial velocity. In this optimal case, one can note that DLL derive
their lowest fluorine abundance, unlike their F-rich star {\it \#III-52} in which the telluric
contribution is much greater and, as a consequence, much more difficult to clean.

\begin{figure*}[t]
\includegraphics[width=8.5cm,height=8cm]{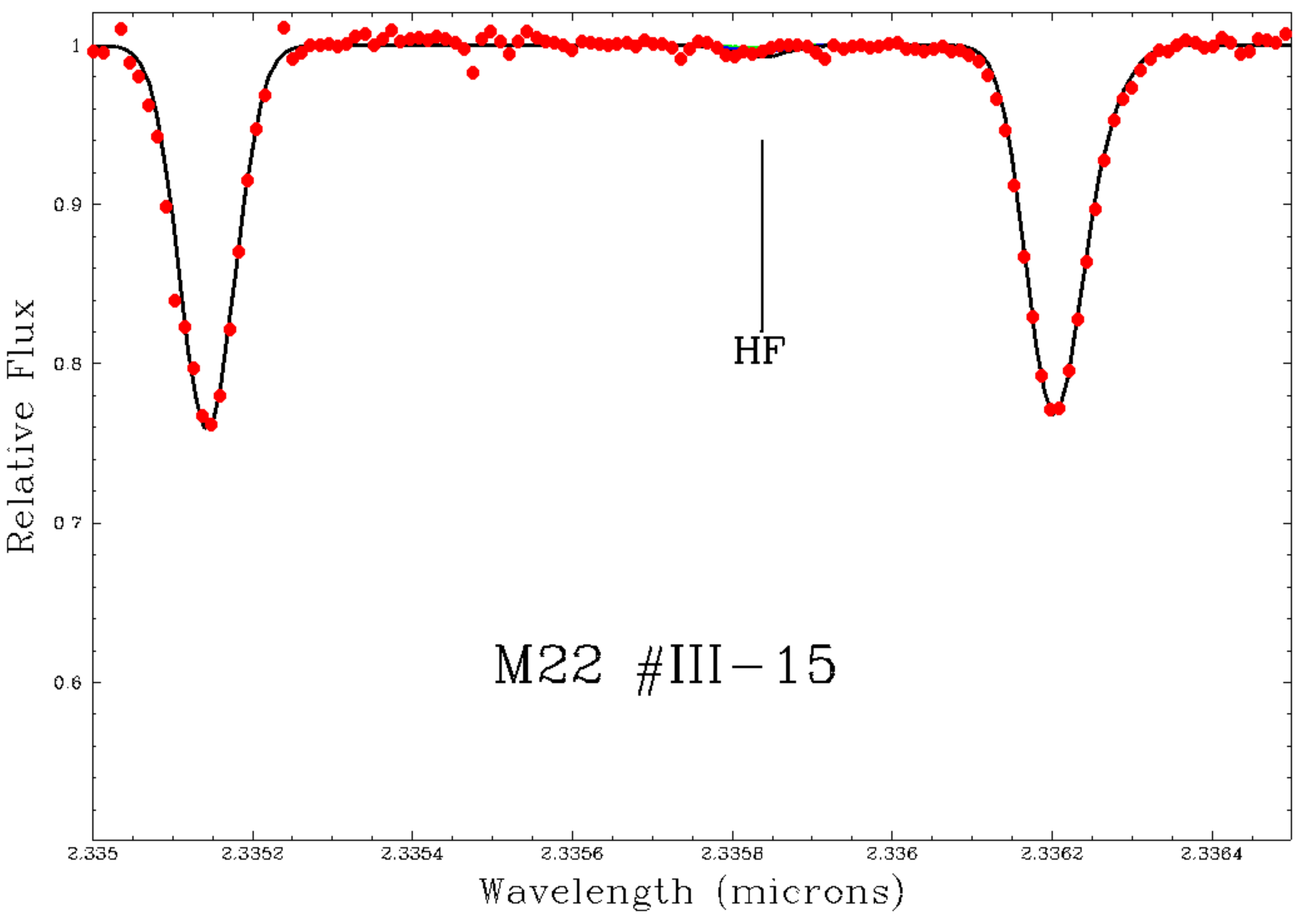} 
\includegraphics[width=8.5cm,height=8cm]{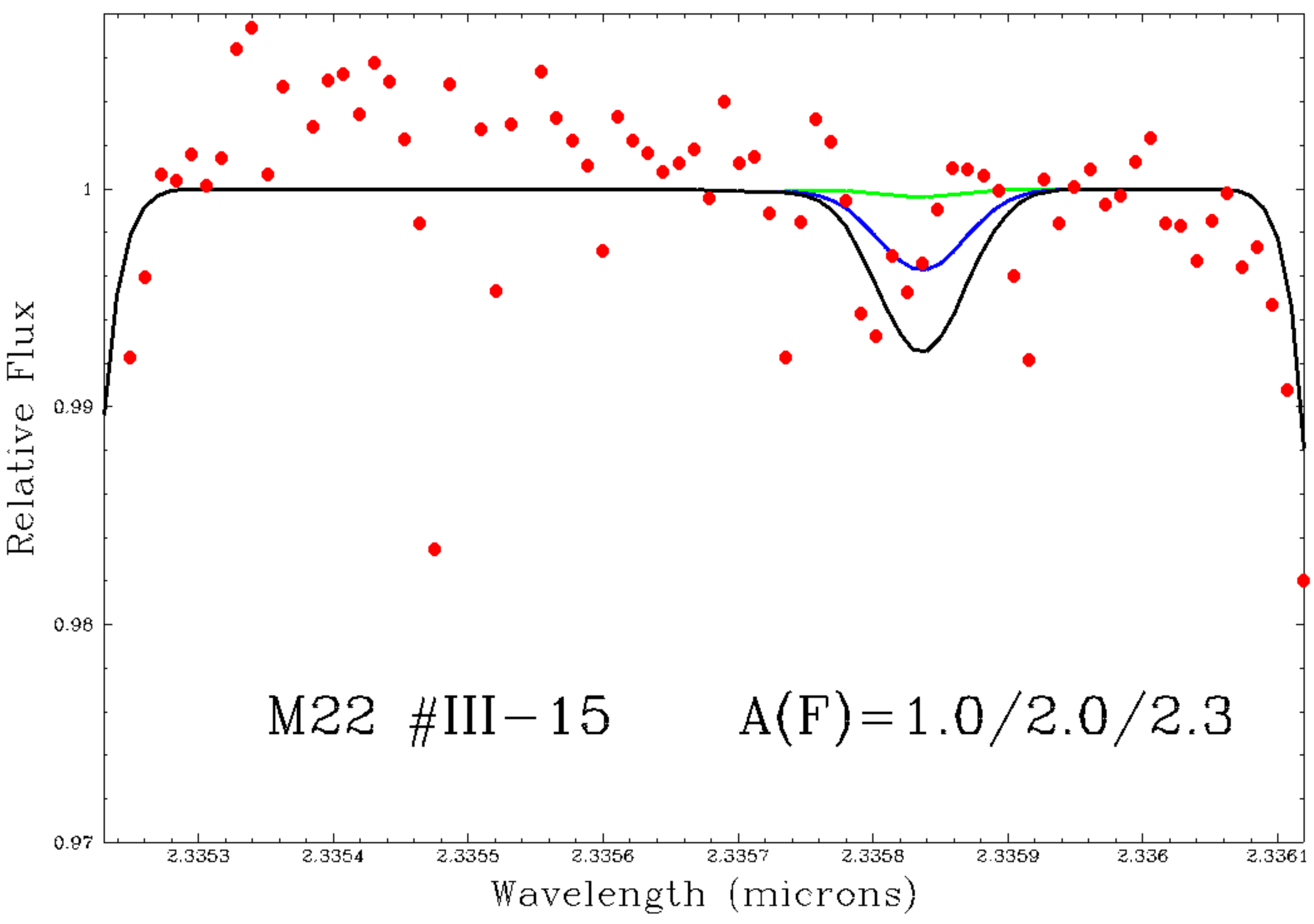}
\includegraphics[width=8.5cm,height=8cm]{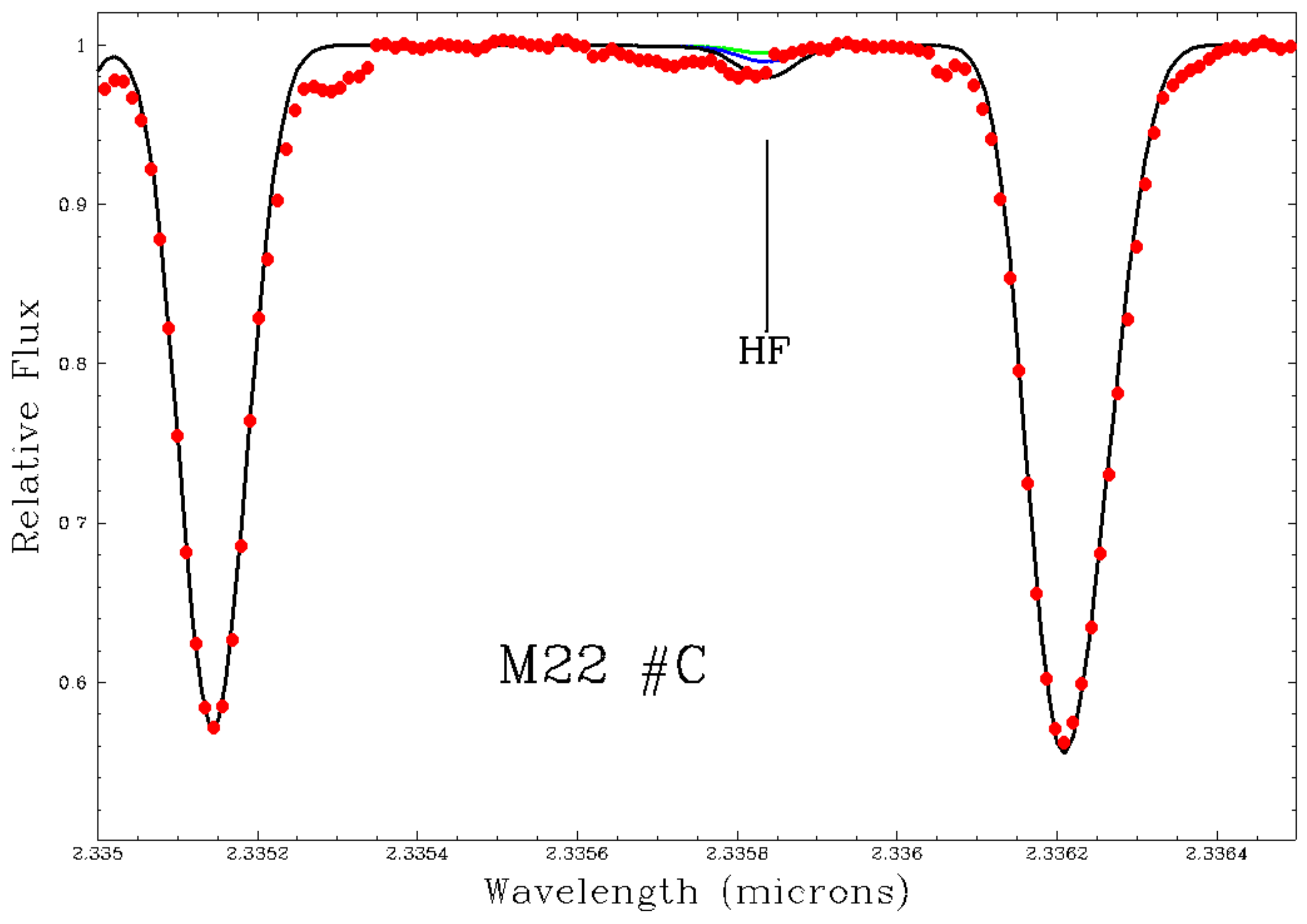} 
\includegraphics[width=8.5cm,height=8cm]{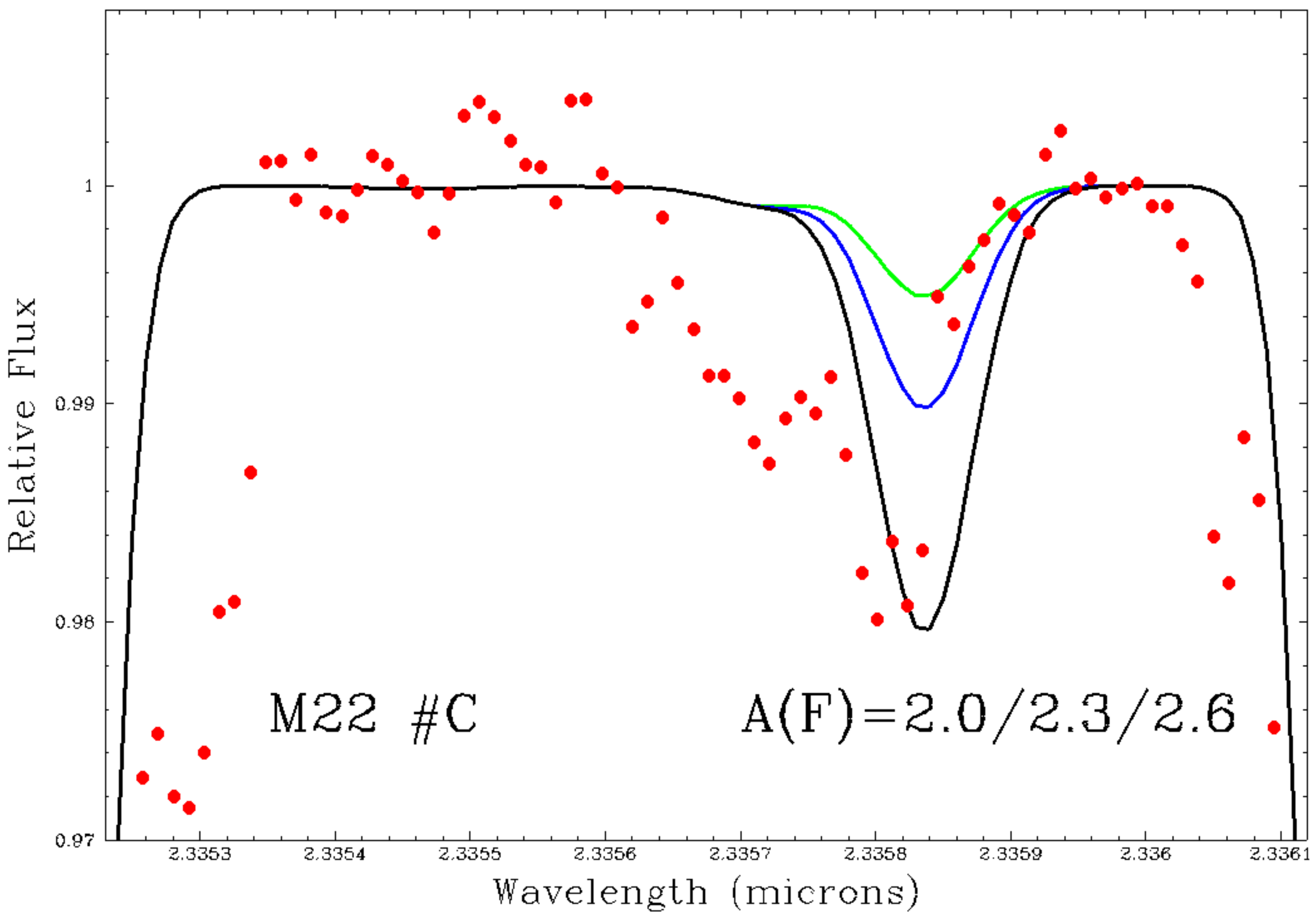} 
\includegraphics[width=8.5cm,height=8cm]{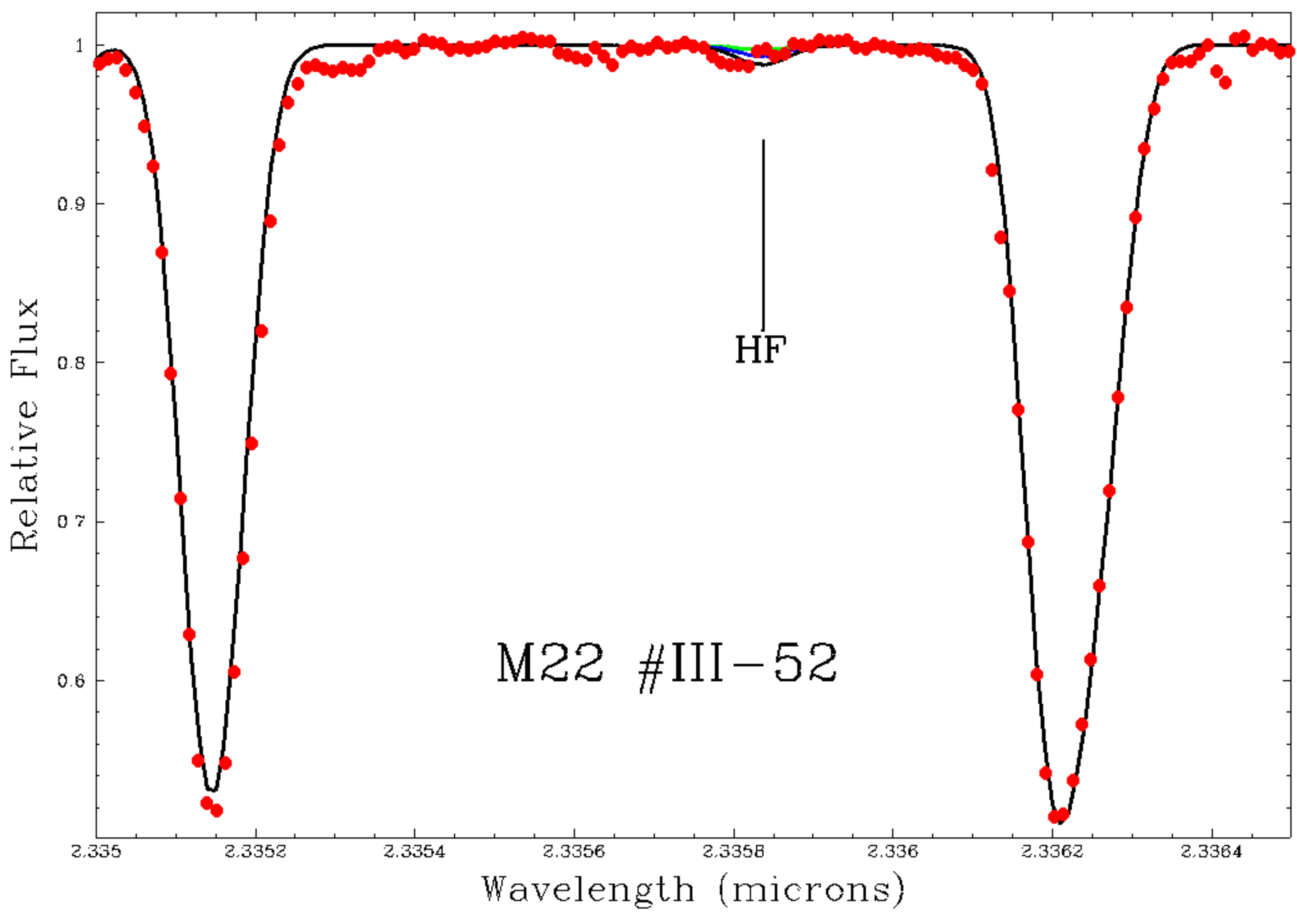} 
\includegraphics[width=8.5cm,height=8cm]{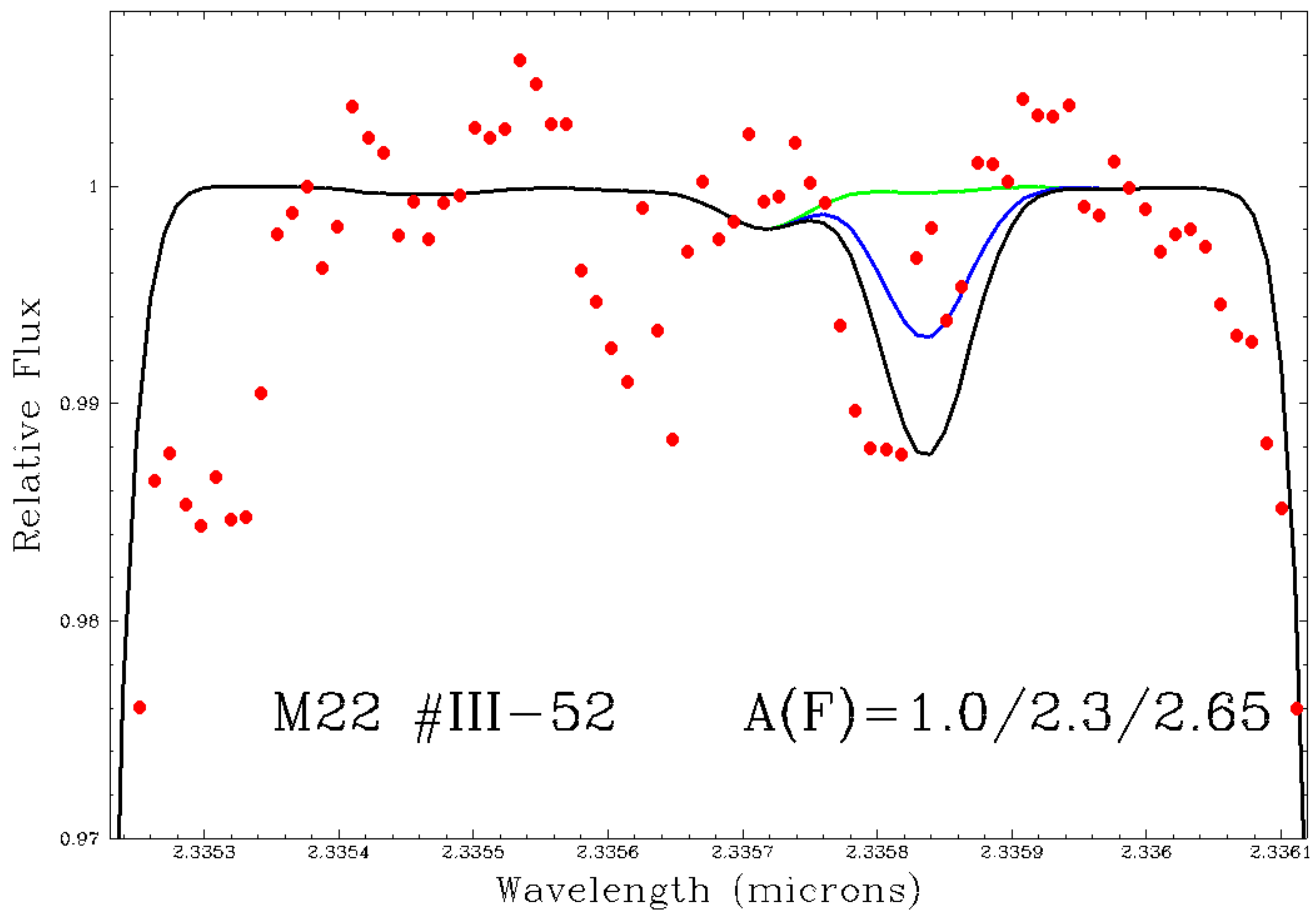}
\caption{Synthetic fits of the observed spectra of the M22 RGB stars {\it \#III-15},
{\it \#C,} and {\it \#III-52} (from top to bottom). The left panels show a rather large spectral domain including the weak HF-transition together with two strong CO lines, confirming the right radial velocity
correction adopted for the fits. A zoom of the left panels around the HF-line 
is shown in the right panels. Three different fluorine abundances spanning a wide range of possible values have been adopted for the synthetic spectra (see the labels in each right-hand panels). }
\end{figure*}

\begin{figure*}[t]
\includegraphics[width=17.cm,height=8cm]{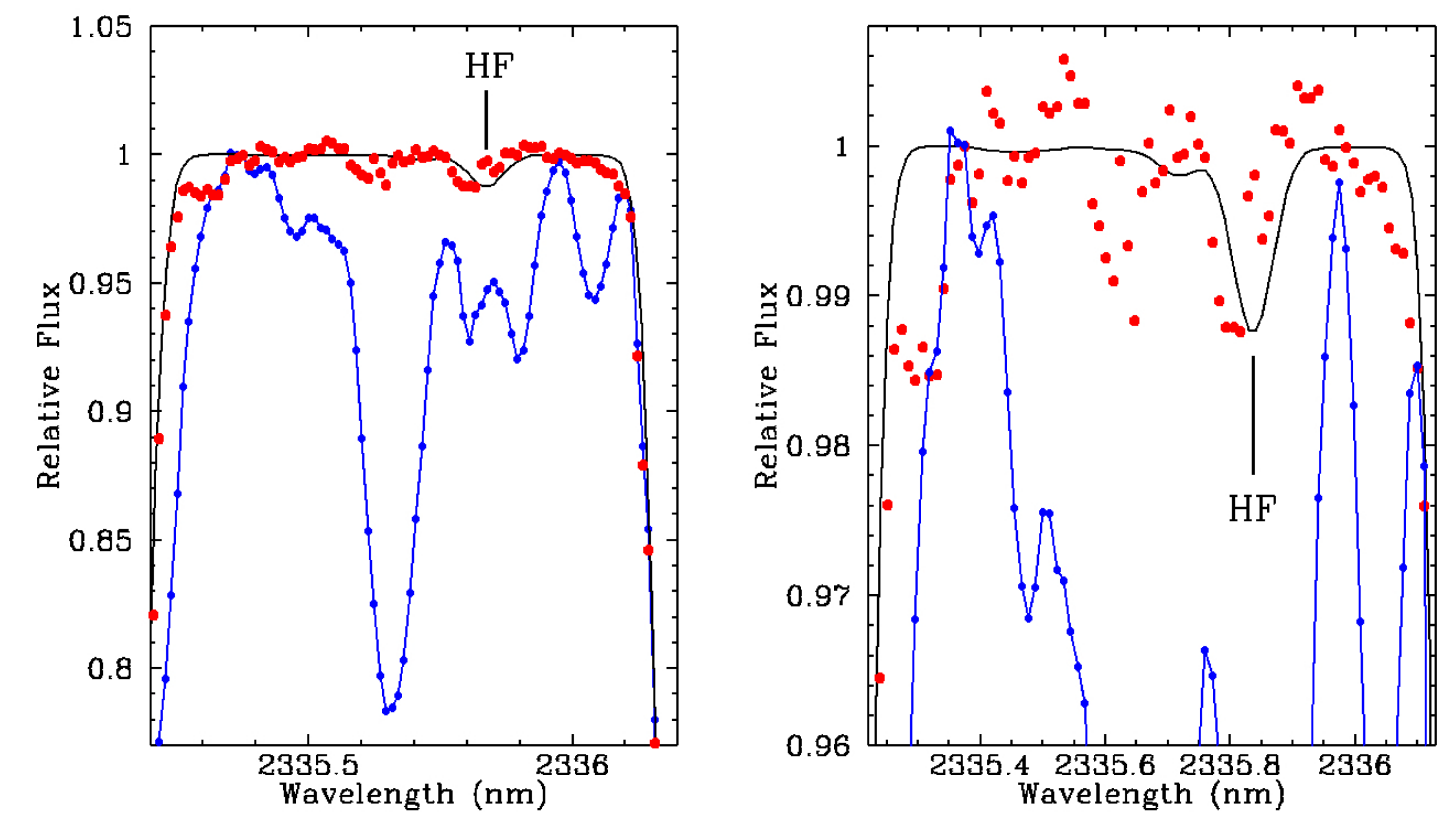} 
\caption{Observed spectrum of the M~22 RGB star {\it \#III-52} with (red points) and without
(blue line and dots) telluric substraction, together with its corresponding synthetic fit (black line).
The cleaned and synthetic {\it \#III-52} spectra are those already shown in Fig.~1 (lower panel).
The left panel is a zoom of the right one around the continuum level where the HF line 
should be found.}
\end{figure*}

\section{Conclusion}
\label{Sect:conclu}
We have re-examined the spectra of three M~22 RGB stars already
studied by DLL13. We showed that the presence of any HF-line
signature is doubtful in these spectra and that, if any of these signatures
were real, only high upper limits or a wide range of possible fluorine contents 
in M~22 RGB stars could be estimated. 
We also proposed that previous estimates of the fluorine
content in this cluster are probably based on incorrect identification
of the HF-transition, owing to 
(i) the presence of continuum fluctuations (probably caused by telluric residuals)
that are as large as the
sought HF line, and (ii) not enough accurate radial velocity corrections.
Indeed, if the radial velocity is determined well using the neighbouring CO lines,
no HF signatures are seen in these spectra,
leading to a totally unknown fluorine content in this cluster.

Therefore, since the three selected targets reanalysed in this note
span the whole range of fluorine abundances derived by DLL13,
it can be concluded that no reliable information on the fluorine content
in these M~22 RGB stars can be derived from the available spectra.
It is thus impossible to claim that there is any fluorine-oxygen correlation
and fluorine-sodium anti-correlation detected in M~22. This avoids any
conclusion on any properties of
the M~22 first-generation stars, such as the range of stellar masses of 
the probable main polluters. 

\begin{acknowledgements}
We acknowledge the anonymous referee for comments
that helped improve this work.
We sincerely thank V. D'Orazi for providing the reduced and cleaned CRIRES spectra
of the M~22 stars reanalysed in this note. R. Gratton is also thanked for his
constructive remarks on the manuscript. 
\end{acknowledgements}

\bibliographystyle{aa.bst}
\bibliography{Fluor_M22}

\end{document}